\documentclass[aps,prl,twocolumn,showpacs,nofootinbib]{revtex4-1}
\usepackage{graphicx}
\usepackage{amsmath}
\usepackage{textcomp}

\begin{document}


\title{Clock hypothesis of relativity theory, maximal acceleration, and M\"ossbauer spectroscopy}

\author{W. Potzel}

\affiliation{Physik-Department E15, Technische Universit\"at M\"unchen, James-Franck-Str. 1, 85748 Garching, Germany}








\date{\today}

\begin{abstract}
Results obtained several years ago using the high-resolution 93.3 keV M\"ossbauer resonance in $^{67}$ZnO and $\beta'$-brass have been reanalyzed with the notion that the clock hypothesis of Special Relativity Theory is not sufficient, but that a maximal acceleration $a_m$ exists and that an acceleration $a$ contributes to the temperature dependence of the center shift by a term $\pm(1/2)(a/a_m)^2$. The significance of the sign of this term is discussed in detail. For both substances a lower limit of $a_m>1.5\cdot10^{21}$m/s$^2$ is inferred which is more than two orders of magnitude \textit{larger} than the value $a_m=1\cdot10^{19}$m/s$^2$ suggested by $^{57}$Fe rotor experiments.
\end{abstract}

\pacs{03.30.+p, 76.80.+y}

\maketitle

\subsection{1. Introduction}
\subsubsection{1.1 Clock hypothesis}

In a M\"ossbauer experiment \cite{Moessbauer} where source (S) and absorber (A) are chemically identical but kept at different temperatures $T_S$ and $T_A$, respectively, the emission and absorption lines of the M\"ossbauer $\gamma$ transition will not have the same energy (frequency). This energy difference is temperature dependent and is caused by a relativistic effect. In a lattice, a nucleus emitting $\gamma$ radiation behaves like a moving clock which exhibits a time dilatation, i.e., a reduction in frequency $\omega$. According to Special Theory of Relativity (STR) \cite{Einstein},\cite{Einsteinbook} we have
\begin{equation}\label{timedil}
t'=\frac{t}{\sqrt{1-v^{2}/c^{2}}}
\end{equation}
thus
\begin{equation}\label{SOD1}
\omega'=\omega\sqrt{1-v^{2}/c^{2}}\approx\omega(1-v^{2}/2c^{2})
\end{equation}
resulting in a frequency reduction $\Delta\omega$ of
\begin{equation}\label{shift}
\Delta\omega=\omega'-\omega\approx-\frac{\omega}{2}\left(\frac{v}{c}\right)^{2},
\end{equation}

where $v$ is the velocity of the moving clock and $c$ is the speed of light in vacuum. This frequency shift \cite{Josephson} is called second-order Doppler shift (SOD), because $\Delta\omega\propto(v/c)^{2}$.
For a M\"ossbauer atom which is vibrating around its equilibrium position in a lattice, $v^2$ is replaced by the mean-square velocity $<v^{2}>$. From equipartition of energy it is estimated that nuclei in a lattice have typical rms velocities of $\sqrt{<v^{2}>}\approx3\cdot10^{2}$ m/s at room temperature. 

Assuming a typical Debye frequency of $\omega_D=10^{13}$ Hz, the rms acceleration of the nuclei is calculated as $\sqrt{<a^{2}>}=\omega_{D}\sqrt{<v^{2}>}\approx10^{15}$ m/s$^2$. Although such accelerations in a lattice are very large, the 'clock hypothesis' \cite{Einstein},\cite{Einsteinbook},\cite{Sherwin} promoted by A. Einstein in the STR asserts they should have no influence on the frequency (energy) of the M\"ossbauer transition. In addition, according to the STR, all velocities $v$ that are allowed relativistically are limited by the condition $v\leq c$. However, there is no limit on the magnitude of acceleration.

\subsubsection{1.2 Maximal Acceleration}
The posssible existence of a maximal acceleration $a_m$ for massive objects was proposed more than 30 years ago \cite{Caianiello}. Based on arguments given in an early paper by Sakharov \cite{Sakharov} that a maximal temperature of thermal radiation should exist, in Ref.\cite{Brandt} a maximal acceleration
\begin{equation}\label{accel}
a_m\approx10^{52} m/s^2
\end{equation}

was suggested. A similar value can be estimated by defining a "Planck acceleration" $c^{2}/l_{P}$, where $l_{P}\approx10^{-35}$m is the Planck length.
The applications and possible consequences of $a_{m}$ cover a huge range of physics, extending from the hyperfine structure of the hydrogen atom and the electron (g-2) anomaly \cite{Fiorentini}, Lamb shift of hydrogen, deuterium and He$^+$ \cite{Lambiase}, violation of Einstein's equivalence principle in connection with neutrino oscillations \cite{Bozza},\cite{Gasperini},\cite{Caianiello1990},\cite{Halprin} to radiation bursts from particles in the field of Compact Impenetrable Astrophysical Objects (CIAOs) \cite{Papini} in relation to black holes. A new dynamics has been introduced in Ref. \cite{Friedman} which extends relativistic dynamics in such a way that a velocity is bound by $c$ and - in addition - an acceleration is bound by $a_m$ \cite{FriedmanResin}, where $c$ and $a_{m}$ are universal constants.

Also the results of M\"ossbauer experiments can be affected if a maximal acceleration exists. This is the subject of the following section.

\subsubsection{1.3 M\"ossbauer spectroscopy: Experiments with high-speed centrifuges using the 14.4keV resonance in $^{57}Fe$}

Several M\"ossbauer experiments were performed with rotors rapidly spinning around a vertical axis, a highly advanced measurement has been reported in Ref. \cite{Kuendig}.  Various geometrical arrangements \cite{WPotzel} have been used: the source being fixed at a certain distance $R_S$ from the center ($R_S=0$, rotational axis) while the absorber is placed at a distance $R_A$, e.g., along the rotor's periphery. The detectors for the M\"ossbauer radiation are stationary located either outside of the rotor or along its vertical axis. With respect to an observer in the laboratory frame, both source and absorber are moving with velocities $\bf{u_{S}}$ and $\bf{u_{A}}$, respectively. If the resonance condition is met, one obtains
\begin{eqnarray}\label{Lee4}
\nu_{A}=\nu_{S}\frac{\sqrt{1-(u_{S}/c)^{2}}}{\sqrt{1-(u_{A}/c)^{2}}}\approx
\nu_{S}\left[1+(u_{A}^{2}-u_{S}^{2})/2c^{2}\right]=
\nonumber\\
=\nu_{S}\left[1+\frac{\Omega^{2}}{2c^{2}}(R_{A}^{2}-R_{S}^{2})\right]
\end{eqnarray}
or
\begin{equation}\label{Lee4a}
\frac{\nu_{A}}{\nu_{S}}=1+\frac{\Omega^{2}}{2c^{2}}(R_{A}^{2}-R_{S}^{2}),
\end{equation}

where $\nu_S$ and $\nu_A$ are the resonance frequencies of the M\"ossbauer $\gamma$ ray as seen in the rest frames of the source (S) and absorber (A), $R_A$ and $R_S$ the radii of the source and absorber orbits, respectively, $\Omega$ is the angular velocity of the rotor. If $R_{A}>R_{S}$, the frequency which the absorber receives from the source is increased, i.e., clocks on a rotating disk run the more slowly, the larger the radius $R$.
If $R_S=0$, the frequency $\nu_S$ of the source has to be \textit{decreased} in order to excite the resonance in the absorber. The relative energy shift is
\begin{equation}\label{Lee4b}
\frac{\Delta E}{E}=-b\frac{\Omega^{2}R_{A}^{2}}{2c^{2}},
\end{equation}

where $b$ is a constant to be determined experimentally. According to SRT, $b=1$.

The technically most sophisticated experiment has been reported in Ref. \cite{Kuendig} where a rotor with a radius of 0.093 m was used and the revolutions per second could be varied between $5 \lesssim \Omega/(2\pi) \lesssim 580$. The result of this experiment was $b=1.0065\pm0.0110$, in perfect agreement with SRT. However, in Ref. \cite{Kholmetskii2008} it has been suggested that an error in the data evaluation of Ref. \cite{Kuendig} occurred and that the value for $b$ should be recalculated to give the corrected result, $b^{cor}=1.19\pm0.03$, which substantially disagrees with SRT. The same experimenters repeated a similar experiment \cite{Kholmetskii2009} and found $b=1.32\pm0.06$, again in serious disagreement with SRT.

From theoretical side, a "new relativistic kinematics of accelerated systems" was developed in Refs. \cite{FriedmanGofman2010}, \cite{Friedman}, \cite{FriedmanarXiv}, \cite{FriedmanResin} with the result that in a rotor experiment where the acceleration and the direction of the M\"ossbauer radiation are along the radius of the disk (longitudinal acceleration), the relative energy shift is determined by the SOD shift plus an additional linear term due to the acceleration $a$:
\begin{equation}\label{linear acc}
\frac{\Delta E}{E}=-\frac{v^{2}}{2c^{2}}-\frac{a}{a_m},
\end{equation}

where $a_m$ is the maximal acceleration. Thus, for a rotor experiment,
\begin{equation}\label{linear acc1}
\frac{\Delta E}{E}=-\frac{\Omega^{2}R_{A}^{2}}{2c^{2}}-\frac{\Omega^{2}R_{A}}{a_m}
=-\frac{\Omega^{2}R_{A}^{2}}{2c^{2}}\left( 1+\frac{2c^{2}}{R_{A}\cdot a_m}\right )
\end{equation}
and
\begin{equation}\label{b-value}
b=1+\frac{2c^{2}}{R_{A}\cdot a_m}
\end{equation}

Using $b=b^{cor}=1.19\pm0.03$, eq. (\ref{b-value}) gives \cite{FriedmanGofman2010}:
\begin{equation}\label{a_{m}-value}
a_{m}=\frac{2c^{2}}{R_{A}(b-1)}=(1.006\pm0.063)\cdot10^{19} m/s^{2}.
\end{equation}

This value for $a_m$ is drastically reduced compared to that estimated in eq. (\ref{accel}).

The acceleration reached in the experiment \cite{Kuendig} was

$a=\Omega^{2} R_{A}=(\Omega R_{A})^{2}/R_{A}\approx350^{2}/0.093\approx1.3\cdot 10^{6}$m/s$^{2}$, thus $a/a_{m}\approx1.3\cdot10^{-13}$, whereas $v^{2}/(2c^{2})\approx6.8\cdot10^{-13}$, assuming that the analysis given in \cite{Kholmetskii2008} is correct. Thus both contributions to eq. (\ref{linear acc}) are significant and of comparable size if $a_{m}$ is given by eq. (\ref{a_{m}-value}).

\subsection{2. Re-evaluation of $^{67}$Zn M\"ossbauer Results}
\subsubsection{2.1 High-resolution spectroscopy and maximal acceleration}

The 93.31 keV transition in $^{67}$Zn has been very attractive due to its extremely high resolution for determining small changes in the $\gamma$-ray energy. The minimal observable linewidth (FWHM) is $2\Gamma_{0}=9.6\cdot10^{-11} eV=0.31\mu m/s\approx 23.4\cdot 10^{3}Hz$. A relative energy resolution of $\sim 10^{-18}$ has been reached. This resonance has been used for precision measurements of lattice-dynamic effects \cite{WPotzel},\cite{WalterPotzel}.

Rotor experiments are characterized by a uniform acceleration between source and absorber. The acceleration $a$ of an atom in a lattice, however, is \textit{not} uniform, in fact the average acceleration $<a>=0$. In analogy to the second-order Doppler shift $(-v^{2}/(2c^{2}))$ and considering eqs. (18) and (19) in Ref. \cite{FriedmanGofman2010} we modify eq. (\ref{linear acc}) to contain the quadratic term $(\pm a^{2}/(2a_m^{2}))$ to take into account a possible additional time dilatation (or enhancement) caused by the acceleration $a$:
\begin{equation}\label{quadratic acc}
\frac{\Delta E}{E}=-\frac{v^{2}}{2c^{2}}\pm\frac{a^{2}}{2a_m^{2}}.
\end{equation}

In the following we will consider data of earlier tem\-perature-shift measurements of $^{67}$ZnO and $\beta'$-brass to derive lower limits for the maximal acceleration $a_m$ and discuss both signs of the acceleration term. At first, we make allowance for the negative sign, in full analogy to the second-order Doppler shift. The positive sign will be discussed in section 3.2.

In M\"ossbauer spectroscopy the temperature dependence of the second-order Doppler shift $(-v^{2}/(2c^{2})$ between source and absorber - within the Debye model - is given by \cite{Wegener}
\begin{eqnarray}
(\Delta E_{SOD}/E)=\frac{u_{SOD}}{c}=
\frac{9k_{B}}{16Mc^{2}}(\theta_{S}-\theta_{A})+
\nonumber\\
+\frac{3k_{B}}{2Mc^{2}}\left[T_{S}\cdot f(T_{S}/\theta_{S})-T_{A}\cdot f(T_{A}/\theta_{A})\right],
\end{eqnarray}

where $u_{SOD}$ is the Doppler velocity in a M\"ossbauer experiment; $T_{S}$, $T_{A}$, and $\theta_{S}$, $\theta_{A}$ are the temperature and Debye temperature $\theta$ of source $S$ and absorber $A$, respectively, $M$ is the mass of the M\"ossbauer nucleus, and $k_B$ is the Boltzmann constant. The Debye integral $f(T/\theta)$ is given by
\begin{equation}\label{Debye integral}
f(T/\theta)=3\left(\frac{T}{\theta}\right)^{3}\cdot\int^{\theta/T}_{0}\frac{x^{3}}{exp(x)-1}dx
\end{equation}


If source and absorber are made of the same material, $\theta_{A}=\theta_{S}$, and only the temperature dependent term is important. In addition, if $T_{S}\ll T_{A}$, only the term characterizing the absorber survives:
\begin{equation}\label{DebyeAbsorber}
\frac{u_{temp}}{c}=-\frac{1}{2}\frac{v_{temp}^{2}}{c^{2}}=-\frac{3k_{B}}{2Mc^{2}}\cdot T_{A}\cdot f(T_{A}/\theta_{A}).
\end{equation}

$u_{temp}$ is the Doppler velocity in a M\"ossbauer experiment where $T_A$ is changed. If $T_{A}\ll T_{S}$, only the term describing the source is important.

\subsubsection{2.2 Measurements on $^{67}$ZnO}

The temperature dependence of lattice-dynamical effects of ZnO single crystals has been investigated between 4.2K and 77.3K \cite{Schafer}. Here only the center shift $S_{C}$ will be discussed. The temperature variation can be written:
\begin{equation}\label{temperature variation}
S_{C}=S_{SOD}+S_{ET}+S_{V}
\end{equation}

The first term describes the second-order Doppler shift (SOD), the second term, $S_{ET}$, the explicit temperature dependence of the isomer shift due to changes of the electron density at the $^{67}$Zn nucleus. The third term, $S_{V}$, represents the volume dependence of the isomer shift (caused by the thermal expansion of the lattice) and was found to be negligibly small: $S_{V}$ amounts to $\sim 0.08\mu$m/s, or only $\sim0.8\%$ of $S_{C}$ observed between 4.2 and 77.3K \cite{WalterPotzel},\cite{Schafer}. $S_{SOD}$ was estimated from specific-heat data \cite{Schafer}. $S_{ET}$ is unexpectedly large, in fact comparable to $S_{SOD}$. $S_{ET}$ exhibits a $T^{4}$-dependence and was found to be relevant already at cryogenic temperatures. Most surprising, $S_{ET}$ shows that the frequency of the $\gamma$-transition in the $^{67}$Zn nucleus is \textit{reduced} if the temperature of the ZnO single crystal is increased. This behaviour was interpreted as a dynamical charge transfer \cite{Shrivastava} of a fraction of a 4s-electron from zinc to the neighboring oxygen atoms \cite{Schafer}. However, in a muon-spin rotation ($\mu$SR) experiment performed later, no evidence for such a charge transfer could be seen \cite{GMK}. Thus the question arises if the observed shift (i.e., a reduction of the $\gamma$-transition frequency) can be explained by the notion of a maximal acceleration $a_{m}$ when the acceleration $a_{temp}$ of the atom is changed due to temperature. If no charge transfer occurs, the sign of the acceleration term in eq.(\ref{quadratic acc}) has to be negative to explain the experimental results. Thus, according to eq.(\ref{quadratic acc}) we have

\begin{equation}\label{tempvari2}
\left (\frac{\Delta E}{E} \right )_{temp}=-\left (\frac{v_{temp}^{2}}{2c^{2}} \right )- \left ( \frac{a_{temp}^{2}}{2a_{m}^{2}} \right ).
\end{equation}

To estimate $a_{m}$ we assume that the Debye model is valid and that 
\begin{equation}\label{tempvari1}
a_{temp}^{2}=\omega_{D}^{2}\cdot v_{temp}^{2}=\left (\frac{k_{B}\cdot \theta_{A}}{\hbar} \right ) ^{2} v_{temp}^{2}
\end{equation}

where $\hbar$ is the Planck constant divided by $(2\pi)$.

Using eqs. (\ref{DebyeAbsorber}) and (\ref{tempvari1}) we obtain for the second term in eq. (\ref{tempvari2}):
\begin{equation}\label{accDebye}
\frac{u_{acc}}{c}=-\frac{3k_{B}^{3}}{2M\hbar^{2}}\cdot \theta_{A}^{2}\cdot T_{A}\cdot f(T_{A}/\theta_{A})\cdot\frac{1}{a_{m}^2}
\end{equation}

where $u_{acc}$ is the Doppler velocity in a M\"ossbauer experiment when acceleration is increased due to temperature.


In the experiments described in \cite{Schafer}, M\"ossbauer spectra were recorded where the temperature of ZnO single crystals was varied between 4.2 and 77.3K. TABLE I summarizes the results obtained in \cite{Schafer} at 40.8, 56.2, and 77.3K where significant changes of $S_{SOD}$ and $S_{ET}$ were observed. The latter we attribute to the term $a_{temp}^{2}/(2a_{m}^{2})$ - in anology to $v_{temp}^{2}/(2c^{2})$ -  and derive a value for $a_{m}$ using eq.(\ref{accDebye}) with $\theta_{A}=275$K and the measured values (due to $S_{ET}$) for $u_{acc}/c$.

\begin{table}
 \begin{center}
 \begin{tabular}{c||@{\hspace{15pt}}c@{\hspace{15pt}}c@{\hspace{15pt}}c}
		\hline
			Temp. & S$_{\text{SOD}}$  & u$_{\text{acc}}\ (S_{ET})$ & a$_\text{m}$ \\
		(K) & (\textmu m/s) & (\textmu m/s) & (m/s$^2$) \\
			\hline
40.8 & -1.01$\pm$0.07 & -0.49$\pm$0.06 & (1.86$\pm$0.13)$\cdot$10$^{22}$ \\
56.2 & -2.5$\pm$0.2 & -1.7$\pm$0.2 & (1.69$\pm$0.11)$\cdot$10$^{22}$ \\
77.3 & -5.5$\pm$0.3 & -3.5$\pm$0.3 & (1.87$\pm$0.08)$\cdot$10$^{22}$ \\
\hline
  \end{tabular}
  \end{center}
 		\caption{Results for ZnO single crystals \cite{Schafer}: Temperature dependences of the second-order Doppler shift $S_{SOD}$ and of the M\"ossbauer Doppler velocity $u_{acc}$ due to $S_{ET}$. The values for the derived maximal acceleration $a_{m}$ are listed in the last column. A negative sign for $S_{SOD}$ and $S_{ET}$ means that the frequency of the $\gamma$-transition in the $^{67}$Zn nucleus is \textit{reduced} if the temperature of the ZnO single crystal is increased from 4.2K to the temperature indicated in the first column.}
 	\label{Tab:Tempdep}
 \end{table}


The errors for $a_{m}$ (see last column of TABLE I) only include the errors of $(u_{acc}/c)$.
The values for $a_{m}$ which agree within these errors are three orders of magnitude larger than the value of $1\cdot10^{19}$m/s$^2$ derived from the rotor experiments \cite{Kholmetskii2008},\cite{Kholmetskii2009},\cite{FriedmanGofman2010}. At low temperatures $(T\ll \theta)$, $f(T/\theta)\approx 3\cdot(\frac{T}{\theta})^{3}$ and $(u_{acc}/c)$ exhibits a $T^{4}$-behaviour (see eq.(\ref{accDebye})), in full agreement with the results of Ref.\cite{Schafer}. Still, on the basis of these results it \textit{must not} be concluded that a maximal acceleration of $a_{m}\approx1.8\cdot10^{22}$m/$s^2$ exists. Only a lower limit for $a_{m}$ can be derived. This aspect will be discussed in detail in section 3. below.

\subsubsection{2.3 Measurements on $\beta'$-brass}

In a similar experiment as for $^{67}$ZnO, lattice dynamical effects have also been investigated for Cu-$^{67}$Zn alloys, in particular for $\beta'$-brass \cite{Peter}. Below $\sim725$K, $\beta'$-brass crystallizes with CsCl structure, where each Zn atom has eight neighboring Cu atoms. This structure is cubic. As a consequence, the M\"ossbauer spectrum depicts a single narrow Lorentzian line. As described in \cite{Peter}, M\"ossbauer spectra were recorded with the $\beta'$-brass absorber heated to various temperatures between 4.2K and 60.0K. For $\beta'$-brass, the phonon frequency distributions have been derived many years ago via phonon dispersion relations obtained from inelastic neutron scattering data \cite{DollingGillat}. On the basis of these neutron data, good agreement was established with the M\"ossbauer results, both concerning the temperature dependencies of the Lamb-M\"ossbauer factor \cite{Wegener} and of the second-order Doppler shift. However, a small shift of $-(0.12\pm0.02)\mu$m/s remained unexplained. If we attribute this shift to the effect of a maximal acceleration we get (using eqs.(\ref{tempvari2}) to (\ref{accDebye}) with $T_{A}=60$K and $\theta_{A}^{\beta'-brass}\approx 250$K): $a_{m}=(7.1\pm1.2)\cdot 10^{22}$m/s$^2$, which is much less accurate but still similar to the values of $a_{m}$ derived above from the $^{67}$ZnO data. In addition, this value is less reliable because $\beta'$-brass  - in contrast to $^{67}$ZnO - exhibits a non-negligible explicit temperature-dependent isomer shift due to lattice expansion at 60K. This additional shift is difficult to estimate with high precision \cite{Adlassnig}.

\subsection{3. Discussion of $^{67}$Zn M\"ossbauer Results}

The values for $a_{m}$ (last column of TABLE I) are based on three aspects: (i) the values for $S_{ET}$ (3rd column of TABLE I) derived in Ref. \cite{Schafer} are valid, (ii) $S_{ET}$ is due to $a_{m}$ according to eq.(\ref{accDebye}) and is not caused by a dynamical charge transfer of a 4s electron, and (iii) the negative sign of the acceleration in eq.(\ref{quadratic acc}) applies. The consequences of all three aspects will now be considered in more detail. In addition, possible corrections in connection with the Debye model will be discussed.

\subsubsection{3.1 Limit for $S_{ET}$.}

As mentioned in section 2.2, $S_{V}$ in eq.(\ref{temperature variation}) is negligibly small. Unfortunately, modern theoretical calculations of $S_{SOD}$ and $S_{ET}$ for ZnO do not exist. To derive reliable experimental values for $S_{SOD}$ and $S_{ET}$ is demanding. However, in Ref. \cite{Schafer} it is shown that $\left|S_{ET}\right|$ has to be within a certain region:

$\left|S_{ET}\right|\ge\left|S_{C}\right| - \left|S_{SOD}^{L}\right|$ and
$\left|S_{ET}\right|\le\left|S_{C}\right| - \left|S_{SOD}^{H}\right|$,

where $S_{SOD}^{L}$ and $S_{SOD}^{H}$ are the second-order Doppler shifts calculated from specific-heat data for low and high temperatures, respectively. $S_{SOD}^{L}$ is only valid for $T\lesssim\theta/10\approx 27K$ and thus not applicable for the temperatures mentioned in TABLE I. The second equation, however, does remain valid, i.e., $\left|u_{acc}\right|=\left|S_{ET}\right|$ has to be smaller than the difference between $\left|S_{C}\right|$ and $\left|S_{SOD}^{H}\right|$. As a consequence, only a \textit{lower} limit for $a_{m}$ can be derived. The results are summarized in TABLE II with the limits $u_{acc}^{lim}$ due to $S_{ET}$ in the 4th column and the lower limits of $a_{m}$ in the last column, assuming a negative sign for the acceleration term in eq.(\ref{quadratic acc}).

\begin{table}[h]
 \begin{center}
 \begin{tabular}{c||@{\hspace{3pt}}c@{\hspace{5pt}}c@{\hspace{5pt}}c@{\hspace{8pt}}c}
		\hline
			Temp. & S$_{\text{C}}$ & S$_{\text{SOD}}^{H}$ & u$_{\text{acc}}^{lim}\ (S_{ET})$ & a$_\text{m}$ (lower limit) \\
				(K) & (\textmu m/s)& (\textmu m/s) & (\textmu m/s) & (m/s$^2$) \\
			\hline
40.8 & -$1.50\pm0.03$ & -0.79 & -0.74 & $1.51\cdot 10^{22}$ \\
56.2 & -$4.22\pm0.04$ & -2.1  & -2.16 & $1.50\cdot 10^{22}$ \\
77.3 & -$9.01\pm0.03$ & -5.0  & -4.04 & $1.74\cdot 10^{22}$ \\
\hline
  \end{tabular}
  \end{center}
	\caption{Temperature dependences of the measured center shift $S_{C}$, of the second-order Doppler shift $S_{SOD}^{H}$ derived from specific-heat data \cite{Schafer}, and of the M\"ossbauer Doppler velocity $u_{acc}^{lim}$ limited by $S_{ET}$. Assuming a negative sign for the acceleration term in eq.(\ref{quadratic acc}), the last column shows the lower limits for the maximal acceleration $a_{m}$ if dynamical charge transfer of 4s electrons is excluded and no corrections in connection with the Debye model are necessary.}
		\label{Tab:Maxacc}
 \end{table}

\subsubsection{3.2 Sign of $\left ( \frac{a^{2}}{2a_{m}^{2}} \right )$ and dynamical charge transfer: Lower limit for $a_{m}$.}

In the following we consider both signs of $\left ( \frac{a^{2}}{2a_{m}^{2}} \right )$, discuss dynamical charge transfer \cite{Shrivastava} when the temperature of ZnO is increased and derive a lower limit for $a_{m}$.

\textit{(a) Negative sign for $\left ( \frac{a^{2}}{2a_{m}^{2}} \right )$}

\textit{No dynamical charge transfer.} Deriving the values for $a_m$ (given in the last columns of TABLE I and TABLE II) we have assumed that $S_{ET}$ is completely caused by $a_{m}$ as described by eq. (\ref{accDebye}) and is not due to a dynamical charge transfer of a 4s electron. This gives a lower limit of $a_{m}>1.5\cdot 10^{22}$m/s$^2$.

\textit{Including dynamical charge transfer.} One could also imagine that the $\mu$SR experiment \cite{GMK} was not sensitive enough to detect such a charge transfer from Zn$\rightarrow$O with high enough precision. If only a fraction $y$ (where $0\le y \le1$) of $S_{ET}$ is due to $a_{m}$ and $(1-y)$ is caused by charge transfer, the left-hand side of eq. (\ref{accDebye}) reads $\frac{u_{acc}}{c}\cdot y$ and $a_{m}$ has to be \textit{increased} by a factor $1/\sqrt{y}$ to fulfil eq. (\ref{accDebye}). In TABLES I and II we have assumed $y=1$; for $y\rightarrow 0$, $a_{m} \rightarrow \infty$. Thus the $^{67}$Zn-M\"ossbauer experiment can only provide a \textit{lower} limit for $a_{m}$.

However, charge transfer might also happen in the opposite direction, from O$\rightarrow$Zn. In fact, a negative shift due to $- \left ( \frac{a^{2}}{2a_{m}^{2}} \right )$ could be largely compensated by a positive shift due to a charge transfer from O$\rightarrow$Zn. To derive a limit for $a_{m}$ we compare ZnO with ZnTe which exhibits the most positive isomer shift (relative to ZnO) of $\sim+98\mu$m$/s^{2}$ (including error bars) of all Zn compounds investigated (s. chapter 4.14 of Ref.\cite{WalterPotzel}). Then taking into account $u_{acc}^{lim}=-4.04\mu$m/s (see TABLE II), $u_{acc}=-(98+4)=-102\mu$m/s. Using eq.(\ref{accDebye}) with $u_{acc}=-102\mu$m/s and $T_{A}=77.3K$, we derive $a_{m}=3.5\cdot 10^{21}$m/s$^{2}$ as a lower limit for $a_{m}$.

\textit{(b) Positive sign for $\left ( \frac{a^{2}}{2a_{m}^{2}} \right )$}

\textit{No dynamical charge transfer.} In this case, the effect of maximal acceleration would cause a positive shift. Although no positive shift is observed, still a \textit{lower} limit for $a_{m}$ can be derived taking the linewidth $\Gamma_{exp}\approx 2.5\mu$m/s$^2$ obtained in the experiments of \cite{Schafer} as a limit for a positive shift. This gives $a_{m}\approx2.2\cdot 10^{22}$m/s$^{2}$.

\textit{Including dynamical charge transfer.} If charge transfer is included a positive shift due to $+\left ( \frac{a^{2}}{2a_{m}^{2}} \right )$ could be compensated by a negative shift due to charge transfer from Zn$\rightarrow$O. Also for this case it is possible to set a limit for $a_{m}$, now by comparing ZnO with ZnF$_{2}$, the latter being characterized by the most negative isomer shift of $\sim-87 \mu$m$/s^{2}$ (including error bars) of all Zn compounds (s. chapter 4.14 of Ref.\cite{WalterPotzel}). Using eq.(\ref{accDebye}), now with positive sign, $u_{acc}=+(87-4)=+83\mu$m/s, and $T_{A}=77.3K$, we obtain $a_{m}=3.8\cdot 10^{21}$m/s$^{2}$ as a lower limit for $a_{m}$, a very similar value as that one derived above for the negative sign.

These lower limits for $a_{m}$ are very conservative, since the extremal values of the isomer shifts (ZnTe and ZnF$_{2}$) were used. This large change in isomer shift is due to the ligands with highly different Pauling electronegativity (s. Fig.30 in Ref.\cite{WalterPotzel}) of 2.1 (Te), 3.0 (O), and 4.0 (F), whereas the change of $S_{ET}$ is due to a charge transfer caused only by a temperature increase from 4.2 to 77.3K. Thus $S_{ET}$ is much smaller than the isomer shifts of ZnTe and ZnF$_{2}$ \cite{Schafer},\cite{Shrivastava}.


\subsubsection{3.3 Debye model.}

ZnO crystallizes with hcp structure and one might dispute if the Debye model is sufficient. However, the $c/a$-ratio of 1.60 is close to the ideal value of 1.633 and the M\"ossbauer data \cite{Schafer} clearly show that - despite the hcp structure - there is very little anisotropy in the recoil-free fraction and that the quadrupole interaction is small. All these data as well as the Gr\"uneisen parameters \cite{Ibach},\cite{Yates} discussed in \cite{Schafer} emphasize that the elastic anisotropy of the ZnO crystal is very small and that the Debye model is able to describe the experimental data quite well indeed.

For ZnO we have used $\theta=275$K. For the derivation of $a_{m}$, the exact value for the Debye temperature is uncritical. For $\theta=300$K, the values for $a_{m}$ in TABLE I and TABLE II would be increased by a negligible factor of $r_{1}<1.03$. This change is so small due to the fact (see eq. (\ref{accDebye})) that the increase in $\theta$ is - to a large extent - compensated by the decrease of the Debye integral $f$.

At low temperatures the Zn and O atoms move together and one could use the sum of the masses of Zn and O. This would mean a reduction of the values for $a_m$ by a factor $r_{2}\sim1.11$.

Within the Debye model, we assumed the equation $a^{2}=\omega_{D}^{2}\cdot v^{2}$. One might argue that $\omega_{D}$ is too large and one should use a lower frequency, $\omega_{D}/r_{3}$. As a consequence, also $a_{m}$ would be reduced to $a_{m}/r_{3}$. A factor of $r_{3}=2$ can be considered as large, because in the Debye model the phonon density of states increases proportionally to the square of the phonon frequency \cite{Walter}.

All three effects might add. Even then, however, $a_{m}$ for ZnO will not be reduced by more than a factor of $r_{2}\cdot r_{3} \sim 2.2$. Thus $a_{m}>(3.5/2.2)\cdot 10^{21}\approx 1.5\cdot10^{21}$m/s$^{2}$ in both cases (ZnO and $\beta'$-brass) which we have considered. A value of $a_{m}=1\cdot10^{19}$m/s$^2$ as derived in \cite{FriedmanGofman2010} is excluded. It would cause shifts which are more than $2\cdot 10^{4}$ times larger (see eq. (\ref{accDebye})) than the Doppler velocities for $u_{acc}$ listed in TABLE I and TABLE II, i.e., they would be in the $cm/s$-range, far outside of the whole observation window of $^{67}$Zn M\"ossbauer spectroscopy. 

Another question which might arise when using the Debye model in connection with the notion of a universal maximal acceleration concerns the motion of a harmonic oscillator in extended relativistic dynamics. As shown in Ref.\cite{Friedman2013}, for small oscillation frequencies $\omega$ the oscillator exhibits the classical behaviour, for large $\omega$ the energy spectrum is similar to that of a quantum harmonic oscillator. Turning to \cite{Friedman2013}, deviations from the classical behaviour start to become noticeable for $\omega\gtrsim7\cdot10^{14}$s$^{-1}$ which, however, is about 20 times higher than the Debye frequency of ZnO $(\omega_{D}\approx 3.6\cdot10^{13}$s$^{-1}$ for $\theta=275$K). In addition, the ZnO measurements were performed at low temperatures (between 4.2 and 77.3K) where a $T^4$-behaviour has been observed as predicted by the Debye model. Thus also regarded from this aspect, the Debye model can be expected to be sufficient for our analysis.

\subsection{4. Conclusions}

An answer to the question if a maximal acceleration $a_m$ exists in Nature is highly important for various aspects of modern physics. The results of previous $^{57}$Fe experiments suggest that $a_{m}\approx1\cdot10^{19}$m/$s^2$. Within the framework of a maximal acceleration we have re-evaluated two measurements obtained earlier in experiments using the high-resolution 93.3 keV M\"ossbauer resonance in $^{67}$ZnO and $\beta'$-brass. The $^{67}$Zn-M\"ossbauer resonance exhibits a minimal observable linewidth of $0.31\mu$m/s which is a factor of $\sim600$ narrower than that of $^{57}$Fe. A lower limit of $a_{m}>1.5\cdot10^{21}$m/$s^2$ is deduced, if the notion of a maximal acceleration is valid. This value is more than two orders of magnitude \textit{larger} than that suggested by $^{57}$Fe rotor experiments.
To reach this limit of $1.5\cdot 10^{21}$m/s$^2$ in future $^{57}$Fe M\"ossbauer experiments will be highly demanding but should also furtheron intensively be examined. We suggest to perform, in addition, synchrotron radiation experiments in combination with high-speed centrifuges at modern synchrotron facilities like PETRA III at DESY (Hamburg) which can cover the relatively high M\"ossbauer energy of the 93.31 keV in $^{67}$Zn. Unfortunately, the resonance in $^{67}$Zn requires to perform experiments at cryogenic temperatures because of the small recoilfree fraction \cite{Wegener}. In addition, as pointed out in Ref.\cite{Friedman2012}, to avoid excessive line broadening great care has to be taken to collimate the beam, thus severely limiting the available count rate. Still, to observe a shift of  $2\cdot\Gamma_{0}\sim0.3\mu$m/s (caused by the linear term $a/a_{m}$ with $a_{m}=1\cdot10^{19}$m/s$^{2}$) would require rotational frequencies $\Omega/(2\pi)$ of only $\sim$50s$^{-1}$ with a radius of the rotor of $0.1$m. However, to reach a lower limit of $a_{m}\gtrsim1.5\cdot10^{21}$m/s$^{2}$, rotational frequences $\Omega/(2\pi)\geq 600$s$^{-1}$ (the maximal value used in Ref.\cite{Kuendig}) are necessary. Thus, for a rotor experiment to become significantly better in sensitivity than the method of using the high accelerations of lattice dynamics described in the present paper will be difficult.

\begin{acknowledgments}
It is a great pleasure to thank A.L. Kholmetskii, Y. Friedman, I. Nowik, G.M. Kalvius, G. Wortmann, R. R\"ohlsberger, A. M\"uller, and F.E. Wagner for stimulating discussions.
\end{acknowledgments}


\end{document}